\pdfoutput=1
\documentclass[twocolumn,superscriptaddress,aps,preprintnumbers,amsmath,amssymb,pral]{revtex4-1}
\usepackage{graphicx}
\usepackage{bm}
\usepackage{color}
\usepackage{amsmath,amssymb}	
\usepackage{hyperref}
\usepackage{setspace}
\usepackage{longtable}
\usepackage{booktabs}
\usepackage{comment}
\usepackage{array}
\usepackage{cancel}
\usepackage{enumitem}
\def\slashchar#1{\setbox0=\hbox{$#1$} 
\dimen0=\wd0 
\setbox1=\hbox{/} \dimen1=\wd1 
\ifdim\dimen0>\dimen1 
\rlap{\hbox to \dimen0{\hfil/\hfil}} 
#1 
\else 
\rlap{\hbox to \dimen1{\hfil$#1$\hfil}} 
/ 
\fi}

\def\beq{\begin{eqnarray}}
\def\eeq{\end{eqnarray}}

\newcommand{\vev}[1]{ \left\langle {#1} \right\rangle }

\begin{document}
\newcolumntype{Y}{>{\centering\arraybackslash}p{23pt}} 


\preprint{IPMU19-0085}

\title{FAKE GUT}

\author{Masahiro Ibe}
\email[e-mail: ]{ibe@icrr.u-tokyo.ac.jp}
\affiliation{ICRR, The University of Tokyo, Kashiwa, Chiba 277-8582, Japan}
\affiliation{Kavli IPMU (WPI), UTIAS, The University of Tokyo, Kashiwa, Chiba 277-8583, Japan}
\author{Satoshi Shirai}
\email[e-mail: ]{satoshi.shirai@ipmu.jp}
\affiliation{Kavli IPMU (WPI), UTIAS, The University of Tokyo, Kashiwa, Chiba 277-8583, Japan}
\author{Motoo Suzuki}
\email[e-mail: ]{m0t@icrr.u-tokyo.ac.jp}
\affiliation{ICRR, The University of Tokyo, Kashiwa, Chiba 277-8582, Japan}
\author{Tsutomu T. Yanagida}
\email[e-mail: ]{tsutomu.tyanagida@ipmu.jp}
\affiliation{T.D.Lee Institute and School of Physics and Astronomy, Shanghai Jiao Tong University, Shanghai 200240, China}
\affiliation{Kavli IPMU (WPI), UTIAS, The University of Tokyo, Kashiwa, Chiba 277-8583, Japan}

\date{\today}
\begin{abstract}
The perfect fit of the matter fields of the Standard Model (SM) into the $SU(5)$ multiplets has strongly supported the idea of the Grand Unified Theory (GUT) for decades.
In this paper, we discuss a novel framework which explains 
why the SM matter fields form the apparently complete $SU(5)$ multiplets.
In the new framework, the apparent matter unification inevitably results from chiral $SU(5)$ gauge theory 
even if the quarks and leptons are not embedded into the common $SU(5)$ multiplets.
We call this class of models the ``fake GUT".
The novel phenomenological prediction of the fake GUT is more variety of the nucleon decay modes than the conventional GUT, which reflects the rich structure of the origin of the matter fields.
\end{abstract}

\maketitle

\section{Introduction}
\label{sec:introduction}
In the Standard Model (SM), the matter fields perfectly fit into three copies of the ${\mathbf {\bar 5}}\oplus \mathbf{10}$  multiplets of $SU(5)$~\cite{Georgi:1974sy}. The renormalization group evolution (RGE) of the three gauge coupling constants of the SM also suggests the unification of the three forces at a high energy scale~\cite{Georgi:1974yf}. For decades, these two aspects of the SM have strongly supported the long-sought idea of the Grand Unified Theory (GUT)  which unifies the three fundamental forces of the SM into a single gauge interaction~\cite{Georgi:1974sy,Georgi:1974yf,Buras:1977yy} (see Ref.~\cite{Tanabashi:2018oca} for reviews). 

The road to the unified theory is, however, not as straightforward as it appears. A closer look at the RGE has revealed that the unification of the three gauge coupling constants is not very precise. The grand unification scale indicated by the RGE also results in too rapid proton decay~\cite{Georgi:1974yf}.

Of course, those apparent failures do not reduce the attractiveness of the GUT. Instead, they may indicate the existence of new charged particles between the electroweak and the GUT scales so that the unification is achieved at higher energy with better precision. The minimal supersymmetric SM (MSSM) is the prime candidate for such a possibility which automatically provides better unification with the higher GUT scale~\cite{Dimopoulos:1981zb,Dimopoulos:1981yj,Langacker:1990jh,Ellis:1990wk,Amaldi:1991cn,Giunti:1991ta,Langacker:1991an}. 
The future confirmation of the MSSM definitely paves the way for the GUT.

In this paper, we discuss another possibility to explain why the SM matter fields form the complete $SU(5)$ multiplets.
As we will argue, the apparently unified matter fields can result even if the quarks and leptons 
are not embedded into the common $SU(5)$ multiplets at high energy scale.
There, chiral $SU(5)$ gauge theory plays a crucial role, though it is only a part of the gauge group 
of the high energy theory.
We call this class of models as the ``fake GUT".
In the fake GUT, no coupling unification is predicted while the matter unification is predicted in a ``fake way".
We will also see that the fake GUT models predict more variety of the nucleon decay modes, which can be tested by ongoing and future experiments.

\section{Fake GUT}
\label{sec:1}
Let us discuss when and how the SM matter fermion fields show up as apparently complete multiplets of $SU(5)$.
For this purpose, it is convenient to consider a quantity $\mit{\Delta} \chi_R(g)\equiv \chi_R(g)-\chi_{R^\dagger}(g)$, where $\chi_R(g)$ is the character (the trace) of a matrix representation $R(g)$ of a group element $g$~\cite{Goodman:1985bw}.
The perfect fit of the SM matter fields into ${\mathbf {\bar 5}}\oplus \mathbf{10}$ of $SU(5)$ means that the sum of $\mit{\Delta} \chi_R$ of the SM matter fields, $A_{\rm SM}$, happens to satisfy
\begin{align}
\label{eq:character}
A_{\rm SM}(g_{\rm SM}) =& n_{\rm gen} \times
\left[
\sum_{i=L_L, \bar{D}_R} \!\!\!\!\!\! \mit{\Delta}\chi_{i}(g_{\rm SM})
+ \!\!\!\!\!\!\!\!\!\! \sum_{j=Q_L, \bar{U}_R, {\bar E}_R} \!\!\!\!\!\!\!\!\!\!\mit{\Delta}\chi_{j}(g_{\rm SM})
\right]\notag\\
=&
n_{\rm gen} \times \left[ \mit{\Delta}\chi_{\mathbf{\bar 5}}(g_{\rm SM})  + \mit{\Delta}\chi_{\mathbf{10}}(g_{\rm SM}) \right].
\end{align}
Here, $\chi_{L_L,{\bar{E}_R},Q_L,\bar{U}_R,\bar{D}_R}$ are the characters of the leptons and quarks, while $\chi_{\mathbf{\bar{5}},\mathbf{10}}$ are 
those of the $SU(5)$ multiplets. 
The number of the generation $n_{\rm gen}$ is three.
The element SM gauge group $G_{\rm SM} (=SU(3)_c\times SU(2)_L\times U(1)_Y) \subset SU(5)$ is denoted by $g_{\rm SM}$.

Since the rank of $G_{\rm SM}$ is the same with that of $SU(5)$ and the characters depend only on the Cartan subgroups, $A_{\rm SM}(g_{SM})$ can be regarded
as a function of the conjugacy class of the full $SU(5)$.
Due to the orthogonality and the completeness of the characters as functions of the conjugacy class of a group element,
the identity Eq.\,\eqref{eq:character} shows that the SM matter fields can be exactly embedded into the three copies of $\mathbf{\bar{5}}\oplus\mathbf{10}$.

The most straightforward way to explain the identity Eq.\,(\ref{eq:character}) is to embed the 
matter fields into ${\mathbf {\bar 5}}\oplus \mathbf{10}$ representations as in the conventional GUT model.
In this case, the GUT predicts
\begin{eqnarray}
\label{eq:characterGUT}
A_{\rm GUT}(g) &=& n_{\rm gen} \times \left[ \mit{\Delta}\chi_{\mathbf{\bar 5}}(g)  +\mit{\Delta} \chi_{\mathbf{10}}(g) \right]  \ .
\end{eqnarray}
Since  $G_{\rm SM}$ has the same rank as $SU(5)$, the fermions contributing to $A_{\rm GUT}(g)$ remain massless in the low energy theory. Thus, the GUT predicts,
\begin{eqnarray}
A_{\rm SM}(g_{\rm SM})  = A_{\rm GUT}(g_{\rm SM}) \ ,
\end{eqnarray}
which explains the identity Eq.\,(\ref{eq:character}).

The identity Eq.\,(\ref{eq:character}), however, can be explained even when the matter multiplets are not embedded into the $SU(5)$ multiplets in the high energy theory.
Let us consider the following conditions of a high energy theory.
\begin{enumerate}
\item The gauge group is $G = SU(5)\times H$, which is spontaneously broken down to $G_{\rm SM}$ completely at the fake GUT scale.
\item Three copies of  ${\mathbf {\bar 5}}\oplus \mathbf{10}$ chiral fermions which are neutral under $H$.
\item  Additional fermions in vector-like representations of $G$.
\item All the Cartan subgroups of $SU(5)$ remain unbroken.
\end{enumerate}
When these conditions are satisfied, the SM matter fields are apparently embedded into 
three copies of $\mathbf{\bar{5}}\oplus\mathbf{10}$ at the low energy,
even if they are not embedded in common $SU(5)$ multiplets at the high energy.

In addition to the above four conditions, we also assume the fifth condition,
\begin{enumerate}[start=5]
   \item At least one of  $SU(3)_c$, $SU(2)_L$, and $U(1)_Y$ is a diagonal subgroup of $SU(5)$ and $H$.
\end{enumerate}
We call the theory which satisfies these five conditions the ``Fake GUT''.

Let us proof that the SM fermions are apparently embedded into 
three copies of $\mathbf{\bar{5}}\oplus\mathbf{10}$ in the fake GUT.
From the conditions, 2, 3 and 4, the sum of $\mit{\Delta}\chi_R$ is again given by
\begin{eqnarray}
\label{eq:characterfakeGUT}
A_{\rm fake\,GUT}(g)&=& n_{\rm gen} \times \left[\mit{\Delta} \chi_{\mathbf{\bar 5}}(g)  +\mit{\Delta} \chi_{\mathbf{10}}(g) \right] \ .
\end{eqnarray}
Note that there are no contributions from the vector-like fermions.
As only the $\bar{\bf 5}\oplus {\bf 10}$ chiral fermions contribute to $A_{\rm fake\,GUT}(g)$, it is a function of $g\in SU(5)$.  
Then, since all the Cartan subgroups of $SU(5)$ remain unbroken,
the fake GUT predicts
\begin{eqnarray}
\label{eq:fakeGUTprediction}
A_{\rm SM}(g_{\rm SM}) = A_{\rm fake\,GUT}(g_{\rm SM}) \ ,
\end{eqnarray}
as in the case of the conventional GUT.
In this way, the identity Eq.\,(\ref{eq:character}) is explained 
in the fake GUT.

It should be emphasized that we have not specified how the SM matter multiplets are embedded in the fake GUT representations.
As we will explain in the following section, some or all of the ${\mathbf {\bar 5}}\oplus \mathbf{10}$ fermions 
may become massive at the fake GUT scale whose mass partners are parts of the vector-like fermions of $G$.
Even in this case, the identity Eq.\,(\ref{eq:character}) is guaranteed by Eq.\,(\ref{eq:fakeGUTprediction}).
Therefore, we find that the apparent unification of the SM matter fields into  ${\mathbf {\bar 5}}\oplus \mathbf{10}$ is inevitably
predicted by the fake GUT where the quarks and leptons are not necessarily embedded into the
common $SU(5)$ multiplets. 
We call this phenomena, the fake matter unification.

As another interesting feature of the fake GUT, the coupling unification is 
not necessarily predicted due to the condition 5.
In fact, when some of the SM gauge groups are diagonal subgroups of $SU(5)$ and $H$,
the corresponding gauge bosons are given by linear combinations of those in $SU(5)$ and $H$.
Accordingly, the gauge coupling constants do not coincide with the $SU(5)$ gauge coupling constant at the 
fake GUT scale.
In general, in the product gauge group models, the three SM gauge couplings are not unified at the fake GUT scale, unless the gauge couplings of $H$ are  not large \cite{Yanagida:1994vq,Weiner:2001pv}.
We will discuss the coupling non-unification in the following section.

The GUT models based on product gauge groups have been discussed in various contexts.
For instance, the models with $H = U(2), U(3)$ have been proposed 
to solve the doublet-triplet mass splitting problem in the SUSY GUT~\cite{Yanagida:1994vq}.
In those applications, however, the matter multiplets are unified into the $SU(5)$ multiplets completely, while the coupling unification is not trivial.
In this sense, the product group GUT  is more close to the idea of the conventional GUT.
In the fake GUT, on the other hand, neither matter multiplets nor the gauge coupling constants are unified in general.

\section{Fake Matter Unification}
\label{sec:2}
As we have mentioned in the previous section, the fake GUT model generally predicts the fake matter unification.
To explain this phenomena, let us first reexamine why the SM matter fields are apparently embedded into 
three copies of $\mathbf{\bar{5}}\oplus\mathbf{10}$ in the fake GUT by using the mass structure explicitly.

Here, let us introduce vector-like multiplets $(\psi,\bar{\psi})$ which are charged under $SU(5)\times H$.
In general, there are $n_i$ fermions, $\bar\psi_i$, whose SM gauge charges are same 
with those of the SM quarks and leptons after spontaneous breaking of $G$.
Here, $i$ runs the species of the SM fermions.
As the $\psi$'s are the vector-like fermions, there are the same number of $\psi$ whose
SM gauge charges are opposite to those of $\bar\psi$.
For $i = L_L$ or $i = {\bar D}_R$, for example, the mass matrix is given by,
\begin{align}
    {\cal L} = \psi_{L_L,{\bar{D}_R}}\, 
    \left({\cal M}^{(L_L,{\bar{D}_R})}_{1}, {\cal M}^{(L_L,{\bar{D}_R})}_{2}\right)
    \left(
    \begin{array}{c}
        \bar{\mathbf 5}_{L_L,{\bar{D}_R}}\\
        \bar{\psi}_{L_L,{\bar{D}_R}} 
    \end{array}
    \right)\ ,
\end{align}
where ${\cal M}_1$ is a $n_i \times 3$ matrix, and ${\cal M}_2$ a $n_i \times n_i$ matrix.
Hereafter, we omit the gauge and flavor indices.
Here, $\bar{\mathbf 5}_{L_L,{\bar{D}_R}}$ denotes the $L_L$ and $\bar{D}_R$ components of $\bar{\mathbf {5}}$,
respectively.
Due to the rank condition of the mass matrix, three linear combinations of $\bar{\mathbf 5}_i$ and $\bar{\psi}_i$  
become massless.
It should be noted that the massless fermions form the complete $\bar{\mathbf 5}$ representation.
For $i=Q_L, \bar{U}_R, \bar{E}_R$, we expect three massless fermions which form the $\mathbf{10}$ representation.

As the GUT symmetry is spontaneously broken, the mass matrices no longer respect the $SU(5)$ invariance, in general.
Therefore, the quarks and leptons have different origins although they form complete $SU(5)$ multiplets,
which is the fake matter unification.

\vspace{.5cm}
{\bf Example}\\
To demonstrate the fake GUT, let us take $H = U(2)$ denoting $U(2)_H=SU(2)_H\times U(1)_H$ 
as an example.

The symmetry breaking, $SU(5)\times U(2)_H\to G_{\rm SM}$, is achieved by the vacuum expectation value 
(VEV) of a complex scalar field, $\Phi^\alpha_i$,
which is a bi-fundamental representation, $({\bf 5},\,{\bf 2})_{-1/2}$, of $(SU(5),\,SU(2)_H)_{U(1)_H}$. 
Here, the subscripts $i$ and $\alpha$ are the indices of the $SU(5)$ and $SU(2)_H$, respectively. The form of the VEV is given by
\begin{align}
\label{eq:vev1}
\vev{\Phi}=\frac{1}{\sqrt 2}\left(
\begin{array}{ccccc}
0&0&0&v&0\\
0&0&0&0&v
\end{array}
\right),
\end{align}
with $v$ being a constant with a mass dimension~\cite{Hotta:1995cd}. 
Once $G$ is broken down to $G_{\rm SM}$,  $SU(3)_c$ appears as an unbroken subgroup of $SU(5)$, 
while $SU(2)_L$ and $U(1)_Y$ appear as diagonal subgroups of $SU(5)$ and $U(2)_H$.

As we discussed in the previous section, the fake GUT contains  three copies of 
${\mathbf {\bar 5}}\oplus \mathbf{10}$ representations of $SU(5)$ and some vector-like representations of $G$ as the matter fermions.
We assume that the vector-like multiplets are given by three pairs of the doublet fermions in $U(2)_H$,
\begin{align}
\label{eq:dl1}
L_{H}: ({\bf 1},~{\bf 2})_{-1/2}\ , \quad \bar L_{H}: ({\bf 1},~{\bf 2})_{+1/2}\ ,
\end{align}
and three pairs of the $SU(2)_H$ singlet fermions,
\begin{align}
\label{eq:dl2}
E_{H}: ({\bf 1},~{\bf 1})_{-1}\ ,\quad \bar E_{H}: ({\bf 1},~{\bf 1})_{+1}\ .
\end{align}

In the presence of the vector-fermions, 
the leptonic components  in the ${\mathbf {\bar 5}}\oplus \mathbf{10}$ multiplets 
can be the mass partners of $\bar{L}_H$'s and $E_H$'s through 
\begin{align}
\label{eq:mixing}
\mathcal{L}=&m_L L_H \bar{L}_H +\lambda_L\bar L_{H}\Phi{\bf  \bar5}\cr 
&+m_E E_H \bar{E}_H+\frac{\lambda_E}{\Lambda_{\rm cut}}  E_{H}{\bf 10}\Phi^\dagger \Phi^{\dagger} \ ,
\end{align}
where $\lambda_{L,E}$ are coupling constants and $\Lambda_{\rm cut}$ denotes a cutoff scale~\cite{Bhattacherjee:2013gr}.
$m_{L}$ and $m_{E}$ are the mass parameters which break the lepton symmetries. 
The cutoff $\Lambda_{\rm cut}$  corresponds to the Planck scale or some heavier particles e.g., scalars with  $(\mathbf{10}, \mathbf{2})$ charge,  which mediate the interaction.
On the other hand, there are no mass partners of the quarks components ${\mathbf {\bar 5}}\oplus \mathbf{10}$,
and hence, the fake matter unification is inevitable in this example.

The leptons in the SM, $L_{L}$ and $\bar{E}_{R}$,  are given by the linear combinations, 
\begin{align}
\label{eq:mix1}
\!\!\!
\left(
\!\!
\begin{array}{c}
L_{M} (\bar E_{M})\\
L_{L} (\bar E_{R})
\end{array}
\!\!
\right)\!\!=\!\!
\left(\!\!\!
\begin{array}{cc}
\cos\theta_{L({E})} & \sin\theta_{L({E})}\\
-\sin\theta_{L({E})} & \cos\theta_{L({E})}\\
\end{array}
\!\!\!
\right)
\!\!
\left(
\!\!
\begin{array}{c}
\bar {\bf 5}_L ( {\bf 10}_{\bar{E}})\\
L_{H} (\bar E_{H})
\end{array}
\!\!
\right).
\end{align} 

Here, ${\bf{\bar 5}}_L$ and ${\bf 10}_{\bar E}$ denote the lepton components of the $SU(5)$  multiplets.
We neglect the flavor dependence of the mixing for simplicity.
In this simple situation, the mixing angles are determined from Eq.\,(\ref{eq:mixing}),
\begin{align}
\tan\theta_L &= \sqrt{2} m_L/(\lambda_L v)\ , \\
\tan\theta_E &= 2 m_E\Lambda_{\rm cut}/(\lambda_E v^2) \ ,
\end{align}
respectively.

As an extreme case, it is possible that $\theta_{L,E} =0$ for $m_{L,E} = 0$.
In this case, the SM leptons completely come from the $U(2)_H$ vector-like fermions, 
while the leptons from  ${\mathbf {\bar 5}}\oplus \mathbf{10}$ become massive.
On the other hand, the quarks come from ${\mathbf {\bar 5}}\oplus \mathbf{10}$,
and hence, the complete fake matter unification is achieved.
This situation can be justified by ``lepton" 
symmetries. 
There, $L_H$'s and $\bar{E}_H$'s are charged, while the others are neutral, which 
makes $m_{L,E} = 0$.

\section{Fake Force Unification}
Next, let us discuss the fake gauge coupling unification.
Due to the condition 5 of the Fake GUT,  at least one massless SM gauge boson is linear combination of the $SU(5)$ and $H$ gauge bosons.
In this case, the corresponding gauge coupling does not coincide with the $SU(5)$ gauge coupling at the fake GUT scale.

{\bf Example}\\
Again we consider the example of $H=U(2)_H$ in the previous section.
With the VEV of the bi-fundamental field $\Phi$ in Eq. \eqref{eq:vev1}, 
the gauge  group $G$ is spontaneously broken to the SM gauge group.

In this example, the SM gauge coupling constants, $g_{1,2,3}$, are given by~\cite{Ibe:2003ys},
\begin{align}
\label{eq:a1}
{1}/{g_1^2}&={1}/{g_5^2}+{3}/{5g_{1H}^2} \ ,\\
{1}/{g_2^2}&={1}/{g_5^2}+{1}/{g_{2H}^2}\ ,\\
{1}/{g_3^2}&={1}/{g_5^2}\ . 
\end{align}
Here, $g_5$, $g_{2H}$, and $g_{1H}$ are the gauge coupling constants of $SU(5)$ and $U(2)_H$, respectively
\footnote{Here, we take the $SU(5)$ normalization for the $U(1)_Y$ gauge coupling.}.
Thus, the moderate unification of the SM gauge coupling constants can be explained 
by choosing appropriate gauge coupling constants of the fake GUT.
In fact, the RGE of the SM gauge coupling constants shows the following relation,
\begin{align}
\label{eq:smrl}
{1}/{g_1^2}\sim{1}/{g_2^2}\gtrsim{1}/{g_3^2}\ ,
\end{align}
at around $10^{14}$\,GeV. 
Thus, the moderate unification can be explained for $ g_{2H} \sim g_{1H}\gg g_5$
when the fake GUT scale is at around $10^{14}$\,GeV.

This extension of the model is partly motivated by the observation that the 
gauge coupling constant $g_{1H}$ could blow up just above the fake GUT scale.
The high-energy theory without the $U(1)$ gauge symmetry is also attractive as it 
explains the charge quantization of the SM straightforwardly.
We will discuss details of this extension in a separate paper. 

\section{Nucleon decay}
In the fake GUT, nucleons can decay through the $SU(5)$ heavy gauge boson $X,Y$ exchange, as the conventional GUTs.
The fake matter unification, however, predicts completely different nucleons decay rates and modes.
For example, when the complete fake matter unification between the SM lepton and quarks occurs, i.e., the SM leptons and quarks have completely different origins, the $SU(5)$ heavy boson mediated nucleon decays are suppressed.
This is also the case with fake matter unification between $Q_L$ and $\bar{U}_R$.
The fake matter unification also leads to more various nucleon decay modes than the conventional GUTs.
Moreover, the additional gauge group $H$ and matters can also contribute to the nucleon decay.
The details of the nucleon decay reflect the gauge and matter structure of the fake GUT.

Let us see the example of $H=U(2)_H$ in the previous section.
In this example, the nucleon decays are dominantly induced by the $SU(5)$ gauge boson exchanges.
For simplicity, we assume that the mixings in Eq.\,(\ref{eq:mix1})  take place generation 
by generation.
In this case, the proton lifetime of the $p\to e^+ \pi^0 $ mode is predicted to be~\cite{Hisano:2012wq,Aoki:2017puj}

\begin{align}
\label{eq:pe}
\tau(p\to e^+ \pi^0 )\! \simeq\! 5\times 10^{26}\sin^{-2}\theta\,\text{yrs}
 \left(\frac{M_X/g_5}{10^{14}\,{\rm GeV}}\right)^4,
\end{align}
for $\theta\equiv\theta_E \simeq \theta_L$.
Here, $M_X$ is the masses of the fake GUT gauge bosons.
The suppression of the decay width by the mixing angle is the generic feature of the nucleon decay in the fake GUT.

The above lifetime is consistent with the current limit~\cite{Miura:2016krn},
$\tau(p\to e^+ \pi^0 )>1.6\times 10^{34}\,\text{yrs}~(90\%\,\text{CL})$, for $\sin\theta\lesssim 10^{-4}$. Such a small mixing angle can be naturally realized when the lepton symmetries are approximately preserved.

It should be noted that the terms in Eq.\,(\ref{eq:mixing}) allow flavor mixings.
Accordingly, the generations of the lepton components appearing in the $SU(5)$ multiplets do not necessarily match with those of the quark components.
Thus, the fake GUT model predicts a variety of the nucleon decay modes.
For example, it  is even possible that the decay rate of the  $p\to  \mu^+ \pi^0 $ mode is larger than that of the $p \to e^+ \pi^0 $ mode.
This is a contrary to the conventional GUT, where the nucleon decay modes which include different generations are suppressed by the Cabibbo-Kobayashi-Maskawa mixing angle.
The fake GUT can explain the two candidate events for $p\to  \mu^+ \pi^0$  in 
Super-Kamiokande IV~\cite{Miura:2016krn} without conflicting the lower limits on lifetime of the $p\to  e^+ \pi^0 $ mode.
Although the observation is not statistically significant at this point, a variety of the nucleon decay modes will provide striking signatures of the fake GUT. 
In this example, the nucleon decay rates and modes reflect not only the gauge structure but also the underling lepton symmetry at the high energy.

\section{Conclusions}
In this paper, we showed that the perfect fit of the SM matter fields into the $SU(5)$ representations 
can be explained by chiral $SU(5)$ gauge theory whether or not the SM matter fields are embedded into the complete multiplets of $SU(5)$.
The ``fake-unification'' of the matter and the force 
is a notable feature of the fake GUT, 
which is never expected in the conventional GUT models.

The fake GUT generalizes the notion of the Grand Unified Theory, which explains 
the matter structure of the SM which miraculously fits into the $SU(5)$ multiplets.
This framework opens up new possibilities of high energy physics 
such as the origin of the flavor structure and the neutrino mass,
where the quark and the lepton flavor structures may obey 
completely different flavor symmetries.
It is even possible to consider the lepton or quark specific gauged flavor 
symmetries, which are not compatible with the conventional GUT.
Those possibilities affect low energy phenomenology.
The detailed study of the nucleon decay reveals the rich matter structures
of the fake GUT.

As a simple example of the fake GUT, we discussed the model based on $SU(5)\times U(2)_H$.
In this model, the SM leptons can dominantly originate from $U(2)_H$ multiplets while the SM quarks  from
$SU(5)$ multiplets in the limit of the lepton symmetries. 
We also showed that the nucleon decay modes which depend on the sizes of the lepton symmetry breaking. 
This prediction can be tested by future nucleon decay searches~\cite{Abe:2018uyc}.
It is tempting to extend the model to $H=SU(3)_H \subset U(2)_H$,
as the $U(1)$ charge quantization of the SM is achieved straightforwardly.

Let us comment that the fake matter unification works even when 
$H$ is trivial, $i.e.$ $G=SU(5),~H=1$. 
For example, by adding vector-like fermions in the minimal $SU(5)$ GUT 
and using the fermion couplings with the ${\bf 24}$ multiplet, 
the quarks and leptons can be embedded into different $SU(5)$ multiplets.
It is even possible to make the nucleon decay rate vanish with fine-tuning of parameters. 
We will discuss the details of this possibility in a separated paper. 
As another example, we may introduce the higher dimensional representation Higgs of $SU(5)$ 
as in the missing partner mechanism~\cite{Grinstein:1982um,Masiero:1982fe}. 
In this case, we can achieve the fake matter unification for $G=SU(5)$ and suppress the nucleon decay rate without the fine-tuning~\cite{Fornal:2017xcj}. 
Those setups are special examples of the present fake GUT framework (see also Refs.~\cite{Hall:2018let,Hall:2019qwx} for $SO(10)$ GUT).

We will explore more general possibilities of the fake GUT in a future work. More general gauge groups and more general matter mixing result in striking features for the nucleon decay. 
It is also interesting to study new possibilities of the flavor symmetries within the framework of Fake GUT, 
such as the $L_{\mu}-L_{\tau}$ gauge symmetry~\cite{Foot:1990mn,He:1991qd,Foot:1994vd} 
with which the observed muon $g-2$ can be explained.

\vspace{-.4cm}  
\begin{acknowledgments}
\vspace{-.3cm}
We thank P.~Cox for useful comments on this paper. We also thank K.~Harigaya for discussion on the importance of the chiral fermion nature for the low energy matter spectrum. 
This work is supported in part by JSPS Grant-in-Aid for Scientific Research No. 16H02176 (T.T.Y), No. 17H02878 (M.I., S.S. and T.T.Y.), No. 15H05889, No. 16H03991 (M.I.), No. 18K13535 (S.S.) and No. 19H04609 (S.S.) and by World Premier International Research Center Initiative (WPI Initiative), MEXT, Japan (M.I., S.S. and T.T.Y.). 
T.T.Y. is supported in part by the China Grant for Talent Scientific Start-Up Project. T.T.Y. thanks to Hamamatsu Photonics.
The work of M.S. is supported in part by a Research Fellowship for Young Scientists from the Japan Society for the Promotion of Science (JSPS).
\end{acknowledgments}

\bibliography{papers}

\end{document}